\documentstyle[12pt]{article}
\begin{document}
\def\Definition #1. {\rm\bigbreak\vskip-\parskip\noindent
       {\bf Definition #1.}\quad}
\def\Theorem #1. {\bigbreak\vskip-\parskip\noindent
{\bf  Theorem #1.} \quad\it}
\def\qed{{\unskip\nobreak\hfil\penalty50\hskip .001pt
  \hbox{}\nobreak\hfil\vrule height 1.2ex width 1.1ex depth -.1ex
  \parfillskip=0pt\finalhyphendemerits=0\medbreak}\rm}

\def\R{{{\rm I} \! {\rm R}}}
\def\O{{\cal O}}
\def\A{{\cal A}}
\def\I{{\cal I}}
\def\B{{\cal B}}
\def\T{{\cal T}}
\def\T{{\cal L}}
\def\E{{\cal E}}
\def\F{{\cal F}}
\def\G{{\cal G}}
\def\Aff{\hbox{Aff}\;}
\def\relation#1_#2^#3{\mathrel{\mathop{\kern 0pt#1}
         \limits_{#2}^{#3}}}
\vspace*{3cm}
\begin{center}
 \Huge\bf
Toposes in General Theory \\
of Relativity
\\

\vspace*{0.35in}

\large

Alexandr K.\ Guts,\ \  Egor B.\ Grinkevich
\vspace*{0.25in}

\normalsize

Department of Mathematics, Omsk State University \\
644077 Omsk-77 RUSSIA
\\
\vspace*{0.5cm}
Email: guts@univer.omsk.su, \quad grinkev@univer.omsk.su  \\
\vspace*{0.5cm}
October 30, 1996\\
\vspace{.5in}
ABSTRACT
\end{center}
We study in this paper different topos-theoretical approaches to the
problem of construction of General Theory of Relativity.
In general case the resulting space-time theory will be
non-classical, different from that of the usual Einstein
theory of space-time. This is
a  new theory of space-time, created in a purely logical manner.
Four possibitities are investigated: axiomatic approach to
causal theory of space-time, the  smooth toposes as a models
of Theory of Relativity, Synthetic Theory of Relativity, and
space-time as Grothendieck topos.

\noindent

\newpage

\setcounter{page}{1}

\section{Introduction}

Construction of a causal theory of space-time is one of the most
attractive tasks of science in the 20th century. From the viewpoint
of
mathematics, partially ordered structures should be considered.
The latter is commonly understood as a set $M$ with a specified
reflexive and transitive binary relation $\preceq$.
A primary notion is actually not that of causality but rather that of
motion (interaction) of material objects. Causality is brought to the
foreground since an observer detects changes of object motion or state.
It is this detection that gives rise to the view of a particular
significance of causes and effects for a phenomenon under study, along
with the conviction that causal connections are non-symmetric.
Causality is treated as such a relation in the material world that
plays a key role in     explaining the topological, metric and all
other
world structures.

Today we imagine  space-time as a world, manifold or {\it set} of
elementary
(atomic) events. An elementary event is a phenomenon whose extension
in
both space and time may be neglected. It is assumed that all phenomena
{\it consist of} elementary events. An event is like a point in
Euclidean
geometry: it is indivisible, or primary. Such an approach allows us to
repeat Euclid's way and to arrive at a geometric theory of space-time.

The manifold of events should represent the material world around us.
The matter exists in no way than in motion that manifests itself in
bodies' influence upon each other. So events also affect each other.
Attempting to follow the process of influence, we simplify the
interaction picture, concentrate out attention on changing states and
thus distinguish causes and effects. So the world appears before our
eyes as a full set of most diverse cause-and-effect connections among
events.

Today we interpreted the world of events as a {\it set}. That means
that the mathematical modelling of the physical space-time was based
on
{\it theory of sets}. This theory has been in the 20th century not
only the
language used by the mathematicians to formulate and realize their
ideas, but also in essence their ideology.
This ideology dictates to us  the necessity of using of the Cantor's
theory of sets
for the construction of mathematical causal theory of space-time
(see, for example, \cite{R, A, Gu, PK, Pi, Pim, Bu}.

Evidently Nature is not
forced to be confined in the sets-theoretic ideological frames of
mathematical abstractions. A transition {\it beyond} the frames of
theory of sets brings new possibilities for describing the real space-time
properties \cite{GT}. An event was so far treated as an indivisible
phenomenon. This is, however, an evident simplification. Time loops,
appearing in general relativity, clearly demonstrate the deficiency of
such an approach. A theory should admit the possibility of automatically
complicating the elementary (atomic) event structure depending on
situation. The structure of (causal) interaction of events should
herewith accordingly complicate, as well as the space-time topological
and metric structures.

Ideally, it would be necessary to have such a formal theory of
space-time that would be able acquire a most sudden appearance relevant
to a concrete model. This approach may be exemplified by obtaining in
\cite{GT} from the same set of axioms, only at the expense of model
(topos) choice, either the flat Minkowski space-time, or the curved
space-time of general relativity.
 ``Topos theory (see ref. \cite{Go, Jo}) offers an
independent (of the set theory) approach to the foundations of
mathematics. Topoi are categories with "set-like", objects,
"function-like" arrows and "Boolean-like" logic algebras.
Handling "sets", and "functions", in a topos may differ from
that in classical mathematics (i.e. the topos {\bf Set} of sets): there
are non-classical versions of mathematics, each with its
non-Boolean version of logic. One possible view on topoi is
this: abstract worlds, universes for mathematical discourse,
"inhabitants" (researchers) of which may use non-Boolean logics
in their reasoning. From this viewpoint the main business of
classical physics is to construct models of the objective
(absolute) universe with a given "bivalent Boolean" model of the
researcher, and choose the most adequate one'' \cite{Tr}.

The topos-theoretical approach to theory of space-time has many
preferences since for one formal space-time theory can exist
physically different models; each topos gives us own physical
world.

Application of topos theory in physics is one of the main
intentions of the creator of this theory W.Lawvere \cite{La}
The idea of using topoi for construction dynamically
variable Universe belongs to russian philosopher I.A.Akchurin
\cite{Ak}. The non-smooth topos theories of space-time was given in
\cite{GT, GD,  GOb}. The variants of smooth topos
theory of space-time
are explained in \cite{GJ,  GGC, GO, Gr, GuI}.
 Application topos theory in quantum theory
can be found in \cite{Tr, Is}.

\section{Toposes and Foundation of Theory \newline
  of Relativity}

The system of axioms for the Special theory of
relativity contains fewer primery notions and relations, is simple,
and lead directly to the ultimate goal (see review \cite{Gu}). In
the case of the General
relativity it is
difficult to introduce a smoothness. This problem was studied by
R.I.Pimenov (see in \cite{Pim} or presentation of
result of \cite{Pim} in \cite{GOb}).

Does the unified way of axiomatization of these
different physical theories exist?
Does the unified way of axiomatization of these
different physical theories exist?
The language of topos theory \cite{Go, Jo} gives the unified way of
axiomatization of
the Special and General Relativity, the axioms being the same
in both cases.
Selecting one or another physical theory amounts to selecting
a concrete topos.

\smallskip

In this section we give a topos-theoretic causal theory of space-time.
Let $\E$  be an elementary topos with an object of natural
numbers, and let $R_{T}$ be the object of continuous real numbers
\cite{St}.

{\it An affine morphism} $\alpha :R_{T} \to  R_{T}$ is a finite
composition of morphisms of the form
$$
1_{R_{T}}, \quad \otimes  \circ  (\lambda \times 1_{R_{T}}) \circ j ,
\quad \oplus  \circ  (1_{R_{T}}\times \mu ) \circ  j ,
$$
where $\oplus ,\otimes $ are the operations of addition and
multiplication in
$R_{T}$ respectively, $\lambda ,\mu $ are arbitrary elements in
$R_{T}$, and $j:R_{T} \simeq 1\times R_{T}$ is an isomorphism.
Let $\Gamma $ be the set of all affine morphisms from $R_{T}$ to
$R_{T}$.

An {\it affine object} in $\E$  is an object $a$ together with two
sets of morphisms:
$$
\Phi  \subset {\rm Hom}_{\E}(R_{T},a), \quad \Psi \subset
{\rm Hom}_{\E}(a, R_{T})
$$
such that the following conditions hold:

1) For any $\phi \in \Phi, \psi \in \Psi $ there is
$ \psi \circ \phi \in  \Gamma $.

2) If $f \in  {\rm Hom} _{\E}(R_{T},a) \setminus \Phi $ then
there
exists $\psi  \in  \Psi $ such that $ \psi \circ f \notin  \Gamma $.

3) If $f \in  {\rm Hom}_{\E}(a,R_{T}) \setminus \Psi $
then there
exists $\phi  \in  \Phi $ such that $ f\circ \phi  \notin  \Gamma$.

4) For any monomorphisms
$f:\Omega \mapsto a , g:\Omega \mapsto R_{T}$
there exists $\phi  \in  \Phi $ such that $\phi \circ g = f$.

5) For any monomorphisms
$f:\Omega \mapsto a, g :\Omega \mapsto R_{T}$
there exists $\psi  \in  \Psi $  such that $\psi \circ f = g $.

Here $\Omega$ is the subobject classifier in $\E$.
\smallskip

An affine object in category ${\bf Set}$ is the set equipped with
an affine structure \cite{Io}. In the topos ${\bf Bn}(M)$ and in the
spatial topos
$ {\bf Top}(M)$ (see notations in \cite{Go}), an affine object
is a fiber bundle
with base $M$ and affines space as fibers.
\smallskip

A categorical description of the Relativity means the introduction
of the
Lorentz structure either in an affine space or in a fiber bundle with
affine spaces as fibers. This can be done by defining in the
affine space a family of equal and parallel elliptic cones or a
relativistic
elliptic conal order \cite{AGS}.

Below we shall use the notations from \cite{Go}.

Let $ a $ be an affine object in the topos $\E $.
\Definition 2.1. {\it An order} in $a$ is
an object $P$ together with a collection of subobjects
$p_{x}:P \mapsto a $, where $x:1 \to  a$ is an arbitrary element,
such that:

1) $x \in  p_{x}$.

2) If $y \in  p_{x}$, then $z \in p_y $ implies $z \in  p_{x}$.
\smallskip
\par\noindent
The order $ \langle P, \{p_{x}\} \rangle $ is denoted as $\O $.

A morphism $f:a \to a $ is called {\it affine}, if
$\psi \circ f\circ \phi  \in \Gamma $ for any $\phi  \in  \Phi $ and
$\psi \in \Psi $. We denote  the set of all affine
morphisms by $\Aff(a)$.

Let $\A \subset  \Aff(a)$ consist of all
commuting morphisms. An order $\O$  is {\it invariant with respect
to \A} if
for any $p_{x},p_{y}$ there exists $g_{xy} \in  \A $ such that
$ g_{xy}\circ p_{x} =p_{y} $.

A morphism $f:a \to  a$ {\it preserves} an order $\O$, if for
each $p_{x}$ there exists  $p_{y}$ such that $ f\circ p_{x} =p_{y} $.
The
collection  of all morphisms preserving an order $\O$  that is
invariant with
respect to $ \A $ is denoted by $ Aut(\O)$.

A {\it ray} is a morphism
$$
\lambda : R_{+} \mapsto R_{T} \relation{\longrightarrow}_{}
^{\varphi}  a,
$$
where
$\phi \in  \Phi_{0} \subset \Phi $, and for any $\phi \in \Phi_{0} $
there is no
$x:1 \to a $ such that $\phi = x \circ !$. Here $ !:R_{T} \to 1 $
and $R_{+}$ is the subobject of object $R_{T}$ consisting of those
  $t$ for which $0 \le t $ (see definition of order in
$R_{T}$ in \cite{St}).

An order $ \O $  is called {\it conic} if 1) for
every $ y \in  p_{x} $ there exists a ray $ \lambda \subset p_{x}$
such that $ x, y \in \lambda $, and 2) $x$ is the origin of
$\lambda$, i.e.
if $ \mu $ is a ray and $ y \in \mu \subset \lambda $,
$ \mu \neq \lambda $,
then $x \notin \mu $.

An order $\O$ has {\it the acute vertex} or {\it pointed} one if for each
$p_{x}$ there does not exist $\phi _{x} \in \Phi_{0} $ such that
$\phi _{x} \subset p_{x}$.  An order $\O$ is {\it complete}, if for any
element $ z:1 \to  a$ and $p_{x}$ there exist  different elements
$ u_{x},v_{x}:1\to a $ and $   \phi \in \Phi_{0} $ such that
$ z,u_{x},v_{x} \in  \phi $ and $ u_{x},v_{x} \in p_{x} $.

An element $u \in  p_{x}$ is called {\it extreme} if there exists
 $\phi  \in  \Phi_{0} $  for which $u \in \phi $, but $ y \notin  \phi $
for all $y \in  p_{x}, y \neq u$.

A conic order $\O$ is said to be {\it strict} if, for each nonextreme
element $u \in p_{x}$, and $v \in p_x, v \neq u$, and each ray
$\lambda$ with
origin $u$ such that $v \in \lambda $, there exists an extreme element
$w \in \lambda $, and $w \in p_x$.

\Definition 2.2. An affine object $a$ with an order $\O$, which is
complete,
strict, conic, has an acute vertex, and is invariant with respect to
$\A$
is said to be {\it Lorentz} if for each $x:1\to a$ and each extreme
elements $u, v \in p_{x}$, where $u, v\neq x$ there exists a
$f \in Aut(\O)$
such that $ f \circ u = v, f \circ x = x $.

\Theorem 2.1. A Lorentz object in the topos {\bf Set} is an affine space
admitting a pseudo-Euclidean structure defined by a quadratic form
$$  x_{0}^{2} - \sum_{i=1}^{n}x_{i}^{2},$$
where $n$ is finite or equal to $\infty $, and $ Aut(\O)$ is the
Poincar\'e
group (see {\rm \cite{AGS}}).  A Lorentz object in the topos ${\bf Top}(M)$
is a
fiber bundle over $M$ with fibers equipped with an affine
structure and a continuous pseudo-Euclidean structure of finite or
infinite
dimension. \rm
\smallskip

 It is quite possible to take not only the topoi ${\bf Set}$,
${\bf Bn}(M)$,
or ${\bf Top}(M)$, but also any others that have an affine object.

The existing categorical determination of the set theory and
determination of ${\bf Top}(M)$ between elementary topoi gives the
possibility to speak about the solution of problem of categorical
description of the Theory of Relativity.
\Theorem 2.2. If $\E$ is a well-pointed topos satisfying the axiom
of
partial transitivity with a Lorentz object $a$, then $\E$ is a model
of set
theory $ Z$ and $a$ is a model of the Special Relativity. If $\E$ is
a topos
defined over {\bf Set} that has enough points and satisfies the
axiom (SG)
(see {\rm \cite{Jo}}) with a Lorentz object $a$, then $\E$ is a topos
${\bf Top}(M)$ and $a$ is a model of the General Relativity. \rm

\section{Smooth toposes as a models of Theory of Relativity}     

There exists  a simple way to construct a non-classical Theory of
Relativity with intuitionistic logic by using
 the objects of smooth toposes
$$
{\bf Set^{L^{op}}, \ {\bf Sh(L)}, \ \G,\ \F}
$$
and some others \cite{MR}.
For this instead of four-dimensional arithmetical space $\R^4$,
adopted in sets theory,
the role of the world of
events passes to its ``analogue''
$$
R^4=\ell C^{\infty}(\R^4)\in {\bf Set^{L^{op}} },
$$
where ${\bf L}$ is the category of loci, i.e. the opposite
category of finitely genereted
$C^{\infty}$-smooth rings.
Every its object $\ell A$ is a $C^{\infty}$-smooth ring which
has the form
$\ell A=\ell C^{\infty}(\R^n)/I$, where
	$I\subset C^{\infty}(\R^n)$
is an ideal and symbol $\ell$ is the label of opposite object.
The category ${\bf L}$  contains the usual category ${\bf M}$
of $C^{\infty}$-manifolds:
$$
   {\bf	M}\subset
{\bf L} \subset {\bf Sh( L)}
	\subset {\bf Set^{L^{op}} }
$$
A morphism or arrow in ${\bf L}$ requires more the explicit
description:

\smallskip
If $B=C^{\infty}(\R^n)/J$, \ $A=C^{\infty}(\R^m)/I$, a morphism
$\ell B\to \ell A$ is an equivalence class of smooth function
$\phi:\R^n \to \R^m$ with property that
$f\in I \Rightarrow f\circ\phi\in J$, while $\phi$ is equivalent
to $\phi^\prime$ if componentwise,
$\phi_i-\phi^\prime_i \in J \ (i=1,...,m).$
\smallskip

Our viewpoint will be to regard   ${\bf Set^{L^{op}} }$
as a generalized set-theoretic universe, where -- intuitively --
every set is a smooth space (and the old Cantor sets are
embeded as discrete spaces).

\smallskip

An event $x$ as {\it an element in the locus stage}
$\ell A = \ell C^{\infty}(\R^n)/I \in \L$
of the space-time $R^4$ is the class of
$C^{\infty}$-smooth vector functions
$(X^0(u), X^1(u),$ $X^2(u),$
$X^3(u)):\R^n\rightarrow  \R^4$, where each function
$X^i(u)$ is taken by $mod \ I$, \  $I$ is a certain ideal of
$C^{\infty}$-smooth functions from $\R^n$ to $\R$.
The argument $u\in\R^n$ is some ``hidden'' parameter corresponding
to the stage $\ell A$. Hence it follows that at the stage of real
numbers $R=\ell C^{\infty}(\R)$ of the topos under consideration
an event $x$ is described by just a
$C^{\infty}$-smooth vector function
$(X^0(u),X^1(u),X^2(u),X^3(u)), u\in \R$.
At the stage of $R^2=\ell C^{\infty}(\R^2)$
an event $x$ is 2-dimensional surface, i.e. a {\it string}.
The classical four numbers
$(x^0,\ x^1,\ x^2,\ x^3)$, the coordinates of the event $x$,
are obtained at the stage
${\bf 1}=\ell C^{\infty}(\R^0)= \ell C^{\infty}(\R)/(t)$
(the ideal $(t)$ allows one to identify functions if their values at
$0$ coincide), i.e., $x^i=X^i(0), i=0,1,2,3$.
\smallskip

The space-time transformations $f:R^4\rightarrow  R^4$ --
are {\it elements at the stage $\ell A$} of the functor
$$
  (R^4)^{R^4}\in {\bf Set^{L^{op}}},
$$
consisting of the classes of $C^{\infty}$-smooth vector functions
 $(F^0(u,x),F^1(u,x),$ $ F^2(u,x)$,
$F^3(u,x)):\R^n\times\R^4\rightarrow \R^4$,
where each function
$F^i(u,x)$
is taken by $mod$ of the ideal
$\pi^*(I)=(\phi\circ\pi\mid\phi\in I, \pi:\R^{n+4}\rightarrow \R^n
- {\rm projection})$. At the stage
${\bf 1}$ these are ordinary transformations without a ``hidden''
multidimensional parameter $u$, while at the stage $R$ these are
smooth transformations with a ``hidden'' parameter.

\smallskip

The relation of causal ordering on $R^4$ can be defined with the
help of the formula
$$
\forall x\in R^4\exists P_x\subset R^4(x\in P_x \ \& \
(\forall y\in P_x
\Rightarrow P_y\subset P_x)\  \&
\ (\forall u\neq v\Rightarrow P_u\neq P_v)),
$$
where  $F\subset G$ means, that a type $F$ is a subtype of type $G$
and one is interpreted in topos ${\bf Set^{L^{op}}}$ as subfunctor
 \cite{MR}. Essentially, at stage $\ell A$  a set $P_x(\ell A)$
consist of classes $C^{\infty}$-smooth vector-functions
$$
((p_x^0(u),p_x^1(u),p_x^2(u), p_x^3(u)):\R^n\rightarrow\R^4)mod \ I;
$$
moreover a term $x$ is interpreted as an element $P_x(\ell A)$, i.e.
class $(X^0(u),X^1(u),$ $X^2(u),$ $X^3(u))mod \ I$. Since $R^4$ at
every stage
can equip the vector structure, i.e. convert in vector space,
then naturally to think, that
  set $P_0(\ell A)$ modulo $I$ is semigroup with respect to addition,
where $0$ at stage $\ell A$ is class $mod \ I$.

The causal automorphisms $f:R^4\rightarrow R^4$ at stage $\ell A$
are defined by relations of the form
$$
F(u, X(u)mod \ I + P_0(\ell A) )mod \ \pi^*(I)
= F(u, X(u)mod \ I)mod \ \pi^*(I) + P_0(\ell A)
$$
for every  $C^{\infty}$-smooth vector function $X(u)$.

It can hope that under some conditions which semigroup
$P_0(\ell A)$ must be satisfied (for example, $P_0(\ell A)$ modulo $I$
is an elliptic cone) the order automorphisms are linear operators,
which at the stage  ${\bf 1}$ coinside with Lorentz
transformations. In any case one appears the new circle of
mathematical and physical problems concerning semigroup theory solving
of which will be useful for given here approach to theory of
space-time...

It is clearly possible to build a causal topos theory
either by analogy with the content of Section 2, or by the scheme used
in Sections 4, 5. However, the resulting space-time theory will be
non-classical, different from that of the Minkowski space-time. This is
a {\it new} theory of space-time, created in a purely logical manner.
It will reflect the real space-time properties to the same extent as
the development of mathematical abstractions accompanies the
development of the real world.

\section{Synthetic Theory of Relativity}

In his work W.Lawvere  \cite{La} suggested a new
approach to the differential geometry and to the others mathematical
disciplines
which are connected with  physics. It  allows to give the definitions
of derivatives, tangent vectors and tangent bundles without
passages to the limits. This approach is based on an idea of consideration
generalized or {\it variable sets} which are the objects of some cartesian
closed category $\E$,
in particular, of some elementary topos.

 The definition of these categories was given by Lawvere and Tirne.
They are possessed of their own inner logic, which in general
are intuitionistic. So it is possible to formulate mathematical
theories with the help of some common logical language and
to give proofs of theorems using the
laws of intuitionistic logic.

The synthetic differential geometry (SDG) is the theory
which was developed by A.Kock \cite{Ko} in the  context of
the Lawrere's ideas.
Basis  of this theory is the assumption  that a geometric line
is not a filled of real numbers, but is  some nondegenerate
commutative
ring $R$ of a line type, i.e such that it satisfies the
following axiom:

\smallskip
{\bf  Kock-Lawvere Axiom }.   {\it Let $D =\{ x\in R \ |\ x^2=0 \}$.
  For every $g:D\rightarrow R$ there exist the unique $a, b\in R$,
  such that for any $ d\in D$  the equation $g(d)=a+d\cdot b$ is valid.}
\smallskip

It follows from this axiom  that all functions
$f:R\rightarrow R$ have the first derivatives.

Further we assume another axioms with respect to $R$ which
allow to state that any $f:R\rightarrow R$ has all derivatives.
The main propositions of classical analysis, for example,
the properties of
derivatives and the Taylor's formula, are valid
for these functions.

In SDG  the original theories of tangent bundles,
differential forms, connections are constructed.  For example,
a tangent vector
to any object $M$ is  a map $t:D\rightarrow M$ and so the tangent
bundle of $M$ is an object $M^D$.

The SDG has several models, and particular, so called
"well adapted models", which allow to compare the classical
differential
geometry with a synthetic one. These models lie in such
categories (toposes) $\E$
that there exist a functors which inset the category ${\bf M}$ of
$C^\infty$-manifolds in $\E$ (see \cite{Ko}). For example,
we can take a smooth topos as the model for SDG.
\smallskip

In our paper we shall define the basic metrical notions in a
context of SDG,
and shall show that it is possible to develop pseudo-Riemannian
geometry  (see \cite{Gr}) and
to write the Einstein's equations.
To this end
 we shall assume that $R$ satisfies to some properties,
which are valid in well adapted models.
\smallskip

We denote as
$InvR=\{x\in R\ | \ \exists \ \ y\in R\ \ x\cdot y=1 \} $
the object of convertible elements in  $R$ and
define an apartness relation on a cartesian product $R^n$ as it follows:
let $x,y \in R^n$, then $x \# y$ iff $\exists i (x_i - y_i \in InvR)$.

We assume that $R$  is a local formally real Pythagorean
Archimedean ring and field of quotients \cite{Ko}.

Having the  apartness relation on $R^n$ we can develop the theory of
intuitionistic linear algebra so as it was made by Heyting
in \cite{He}.
In particular,  the intuitionistic  theory of linear
equations is valid.
The notion of basis of $R$-module are also defined.
\smallskip

Having the apparat of linear algebra and using the assumptions
with respect to $R$ that were  given above we  define a scalar
product on $R^n$
and show that  the main metrical properties of $R^n$
do not differ from classical one. For example,
if the determinant of matrix of a scalar product is aparted
of zero than the matrix  is invertible.

For $R$-modules $U$ and $V$ we define a tensor product
$U\otimes V$  and all operations of tensor algebra in
a standard manner.
Since $R^n$ is a $R$-module we can define the
notions of covariant and contrvariant tensors on $R^n$, and by using
the properties of derivatives and the Taylor's formula we show
the existence of analogy of classic tensor analysis. Having a
definition of scalar product we define
the operation of raising and lowering of indexes.

\smallskip

The next result is the definition a Riemannian tensor on a
formal manifold. {\it A formal manifold} is a notion of SDG
which is represent a classical notion of manifold \cite{Ko}.
It has a number of good and natural properties. It is possible
to define  a local charts of points which are formal etale
subojects of $R^n$, and to show that tangent $T_pM$ space at each
point $p$ has the structure of $R$-module and is isomorphic
to $R^n$.
\smallskip

  Let $M$ be a formal manifold.
  A map $g:TM\times_M TM\rightarrow R$
  is called {\it pseudo-Riemannian metric or structure }
  on $M$ if the following conditions are satisfied:
  \begin{enumerate}
   \item  $v\# 0 \Rightarrow \exists u: g(v, u)\# 0 $\\
          $v=0   \Rightarrow g(v, v)=0 $
    \item $g(v, w)=g(w, v) $
    \item $g(u+v, w)=g(u, w)+g(v, w) $
    \item $g(\lambda \cdot v, w)=\lambda \cdot g (v, w) $
  \end{enumerate}
  where $\lambda \in R$, $v, w, u \in TM $ so that $v(0)=w(0)=u(0)$.
\smallskip

A map $g_p:T_{p}M\times T_{p}M\rightarrow R $ is
a scalar product on $T_{p}M$ if
$g_{p}(u,v)=g(u,v)$  for $u, v\in T_{p}M$,

 For any $u, v\in T_{p}M $ we have
$g_p (u, v)=(g_p)_{ij}\cdot u^i v^j,$
where $ u^i, v^j$ are coordinates of vectors $u, v$
in  the basis  $\{ \partial_i \}$
and $(g_p)_{ij}=g_p(\partial_i ,\partial_j)$.
We have that $det\|(g_p)_{ij}\|\# 0 $ and matrix
$\|(g_p)_{ij}\|$ is  invertible.

By using the fact that tangent spaces
are $R$-modules
we  define the tensor bundles under formal manifold.

In  \cite{KoR} was developed the original theory of connections
in a context of SDG. There it was shown that connections on the
bundles over a subobject $U$ of $R^n$ were defined by $3n$ coefficients
$\Gamma_{ij}^k$  and that the tensor of curvature
(tensor of Riemann-Christoffel) had the classical
expression in coordinates.

So we define the connection on formal manifold
by means of definition of  $\Gamma_{ij}^k$ in a local chart
and than introduce the tensor of curvature by using
the its coordinates $R_{ijk}^l$ in a local chart
which are expressed in a classical manner.
It is evident that in this case the Riemann-Christoffel's
tensor  has a standart properties.
Further we can define the Ricci's and Einstein's tensors by
using the classical operations over tensors.

Now  we can write the Einstein's
equations of gravitational field if we have a four-dimensional
formal manifold with
a given pseudo-Riemannian structure.

There exist the models of pseudo-Riemannian structure on a formal
manifold in different well adapted  models.

So we have the  method of construction of models of intuitionistic
General Theory of Relativity in cartesian closed categories,
and, in particullary, in toposes.

In model ${\bf Set^{L^{op}}}$ the Einstein's equations at stage
$\ell A =\ell C^\infty(\R^n)/I$ in vacuum have the following form:
$$
   R_{ij}(u) - \frac{1}{2} g_{ij}(u) R(u) = 0 mod I,  \eqno(*)
$$
where the argument $u\in\R^n$ is some ``hidden'' parameter
corresponding
to the stage $\ell A$, i.e. we got the non-denumerably infinite
collection of equations.
The solution of these equations is
serious problem.
At stage
${\bf 1}=\ell  C^\infty(\R^0) = \ell C^\infty(\R)/(x)$
the equations (*) coinside with the usual Einstein's equations.

\section{Space-time as Grothendieck topos}

There is still another possibility of applying topos theory to
a mathematical description of space-time. One can attempt to
achieve the desired simplicity when axiomatizing relativity theory
at the cost of giving up the classical view that space-time
is the world of events "placed" in a single "space".
\smallskip

To this
end, consider a partially ordered set $<{\bf P}, \preceq >$
and contravariant
functors from the pre-category ${\bf P}$ to the topos {\bf Set}. This
gives rise to the topos ${\bf Set^P}$, and it is this topos which
is the {\it new mathematical space-time}.
\smallskip

The value of a functor $F$
on an element $x$ of ${\bf P}$ is the set $F(x)$. The set ${\bf P}$ is
interpreted as the collection of all possible situations of
obtaining information about past. It has a  temporal partial
order. The set $F(x)$ is the (causal) past cone consisting
of the events that are observed in situation $x$. The
functor $F$ can be interpreted as a time flow. The topos
${\bf Set^P}$ consists of all possible time flows. It is
not hard to see that a classical Lorentz transformation
corresponds to a natural isomorphism of functors, i.e.
time flows.
  In fact, consider two different  time flows $F$ and $G$, and let
$x,y, \ \  x \preceq y$ are two time situations. Then we have the
following diagram:

\begin{eqnarray*}
     x \quad    & \preceq         &  \quad y    \\
                &                 &                \\
      F(x)      & \stackrel{F(\preceq)}{\longrightarrow} & F(y) \\
\downarrow \tau_x  &          & \downarrow \tau_y   \\
      G(x)& \stackrel{G(\preceq) }{\longrightarrow} & G(y)
\end{eqnarray*}
where $\tau$  is a natural isomorphism of functors $F, G$.
Let all $F(x), G(x), \ x\in {\bf P}$ are "placed" in a single space
$\R^4$. Assume that $F(x)$ is an elliptic cones with vertex $x$;
 \ $F(x)$ is equal and parallel to $F(y)$,  $G(x)$ is equal
and parallel to $G(y)$
 and $\tau_x=\tau : \R^4\to \R^4$ for all $x\in {\bf P}$
is a mapping such that
$$
		 \tau(F(x))=G(x),
$$
$$
G(x)=F(\tau(x)) \eqno(**).
$$
 As it follows from \cite{A, Gu} the mapping $\tau$ is
a classical Lorentz transformation. But if the relation (**)
is not valid, for example, $G(x)=\phi_x (F(x))$, where $\phi_x$
is a rotation on constant angle with respect to point $x$,
\ \ $G(x)\neq F(x)$, i.e.
we have the flow of time (life) there where we see the present (death),
 then the theory of sets is useless.

\smallskip

Thus, the space-time ${\bf Set^P}$, which may be described as
a Grothendieck topos \cite{Go}, can no longer be "placed" in a single
 "space".

\end{document}